\documentclass[twocolumn,showpacs,preprintnumbers,amsmath,amssymb]{revtex4}

\usepackage{graphicx}
\usepackage{dcolumn}
\usepackage{bm}
\usepackage{subfigure}

\begin{document}

\title{Numerical blowup in two-dimensional Boussinesq equations}

\author{Z. Yin}
\email{zhaohua.yin@imech.ac.cn}

\affiliation{National Microgravity Laboratory, Institute of Mechanics, Chinese Academy of Sciences, Beijing 100190,
P.R.China}

\date{\today}

\begin{abstract}

In this paper, we perform a three-stage numerical relay to investigate the finite time singularity in the
two-dimensional Boussinesq approximation equations. The initial asymmetric condition is the middle-stage output of a
$2048^2$ run, the highest resolution in our study is $40960^2$, and some signals of numerical blowup are observed.

\end{abstract}


\maketitle

It is still an open question whether smooth initial conditions in three-dimensional (3D) Euler equations can develop
singularities in a finite time. The current numerical methods can not really express an infinite number in a dynamic
simulation, so besides the right initial conditions, extremely high resolutions are necessary to catch the signal of
blowup. The obvious ways to increase the resolution are: 1) Using the largest computer available to perform
simulations; 2) Adopting some kinds of symmetries in the 3D Euler equations~\cite{tay35,bra83,kid85}. The
two-dimensional (2D) Boussinesq approximation equations correspond to the 3D axisymmetric Euler equations, and need
much less computer capacity than those 3D symmetric models. In the meanwhile, it can reveal much more physics than the
one-dimensional symmetric model~\cite{yin2006,yin2005}. The blowup signal of 2D Boussinesq equations is derived in
\cite{e94,mar02}: if the maximum absolute values of the vorticity and temperature gradient behave like
\begin{equation}
(T_c - t)^{-\alpha} \; \& \; (T_c - t)^{-\beta}  \label{eqcc}
\end{equation}
with $\alpha > 1$ and $\beta > 2$, a finite time singularity will be developed.

The equations under consideration are the following~\cite{mar02}:
\begin{eqnarray}
&& \theta_t + {\rm {\bf u}} \cdot \nabla \theta = 0, \label{eq23}
\\
&& \omega _t + {\rm {\bf u}} \cdot \nabla \omega = - \theta_x, \label{eq24}
\\
&& \Delta \psi = - \omega, \label{eq25}
\end{eqnarray}
where $\theta$ is the temperature, ${\rm {\bf u}} = (\mbox{u,v})$ the velocity, ${\rm { \mbox{ \boldmath{$\omega$}}}} =
(0,0,\omega ) = \nabla \times {\rm {\bf u}}$ vorticity, and $\psi$ stream function.

The method adopted in our numerical simulations is the filter pseudo-spectral method with some proper de-aliasing
technique. The modifying factor for each Fourier mode $k$ in the filter is $\varphi(k) = e^{-37(2k/N)^{16}}$ for $k <
N/2$, where $N$ is the grid number in one direction. The de-aliasing scheme to be used is the phase-shift scheme with
circular truncation ~\cite{pat71}.

The low-storage third-order Runge-Kutta time discretization is adopted. Besides its high accuracy, the one-step
property of this scheme is essential to our three-stage numerical relay~\cite{can87}.

During the process towards the singularity, a $\delta$ or $\delta$-like function will be developed in the flow field.
It is well known that spectral coefficients of the $\delta$ function are constant for different modes:
\begin{equation}
\delta(x,y) \simeq C \sum^{\frac{N}{2}}_{n = -\frac{N}{2}} \; \sum^{\frac{N}{2}}_{m = -\frac{N}{2}} e^{-i(nx + my)},
\label{eqrr}
\end{equation}
where $C$ is a constant. If the resolution $N^2 \rightarrow \infty$, we will get the exact Fourier representation,
which is impossible in current computers. The main idea of this research is to make $N$ as large as possible to capture
the blowup signal.

For high-resolution simulations, it is very important to have an effective initial condition because a poorly chosen
initial condition may not lead to blowup or may lead to blowup after an unacceptable long-time computation. In the
following, we will describe how to obtain the initial data for the whole paper. First, we take the initial condition
with unified zero vorticity and a cap-like contour of temperature with the following expression:
\begin{equation}
\theta (x,y,0) = 50(\frac{4x-3\pi}{\pi}) \theta _1 (x,y)\theta _2 (x,y)\left[ {1 - \theta _1 (x,y)} \right],
\label{eq35}
\end{equation}
 where if $S(x, y):=\pi^2 - y^2 - (x - \pi )^2$ is positive, $\theta_1 = \exp {\left( 1
- \pi^2/S(x,y)\right)}$, and zero otherwise; if $s(y):= \left| {y - 2\pi } \right| /1.95\pi$ is less than 1, $\theta_2
= \exp \left( 1- (1- s(y)^2) ^{-1} \right)$, and zero otherwise. We compress the intermediate results (at $t=1.2$)
starting from the above initial condition to form a new initial data. More precisely, we let $\omega(x,y,0) =
\omega'(x,2y-0.4\pi,1.2)$ and $\theta(x,y,0) = \theta'(x,2y-0.4\pi,1.2)$, for $(x, y)\in [0, 2\pi] \times [0, \pi]$
(where $\theta'$ and $\omega'$ are obtained by solving Eqs. (\ref{eq23})-(\ref{eq25}) and(\ref{eq35})  with a $2048^2$
grid), and zero otherwise. To eliminate the high order frequency generated from the compression, we perform a $2048^2$
run with the new initial data. The intermediate results at $t = 0.12$ are the REAL initial data for the whole paper.
The flow field in spectral space is stored, with 2048 modes in both directions.

The system describes a cap-like hot zone of fluid rising from the bottom, while the edges of the cap lag behind,
forming eye-like vortices (Fig. (\ref{fig:contour})). The hot liquid is driven by the buoyancy and meanwhile attracted
by the vortices, which leads to the singularity-forming mechanism in our simulation. The flow filed is asymmetric to
$x=\pi$ to avoid other mechanism during the singularity forming~\cite{yin2006,yin2005}. Three-stage simulations are
planned:

\begin{figure}
\begin{minipage}[c]{0.8 \linewidth}
\scalebox{1}[1]{\includegraphics[width=\linewidth]{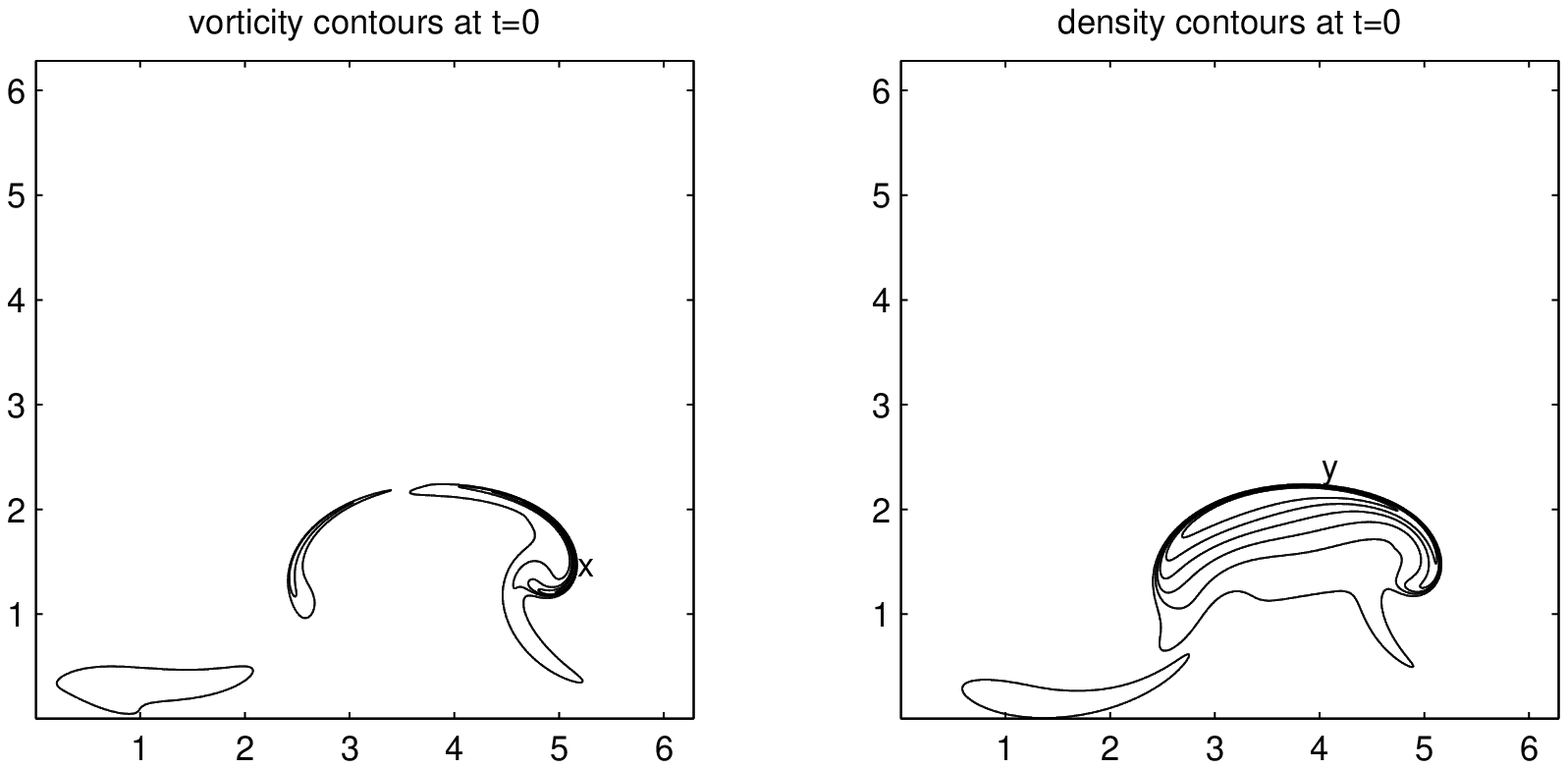}}
\end{minipage}
\begin{minipage}[c]{0.8 \linewidth}
\scalebox{1}[1]{\includegraphics[width=\linewidth]{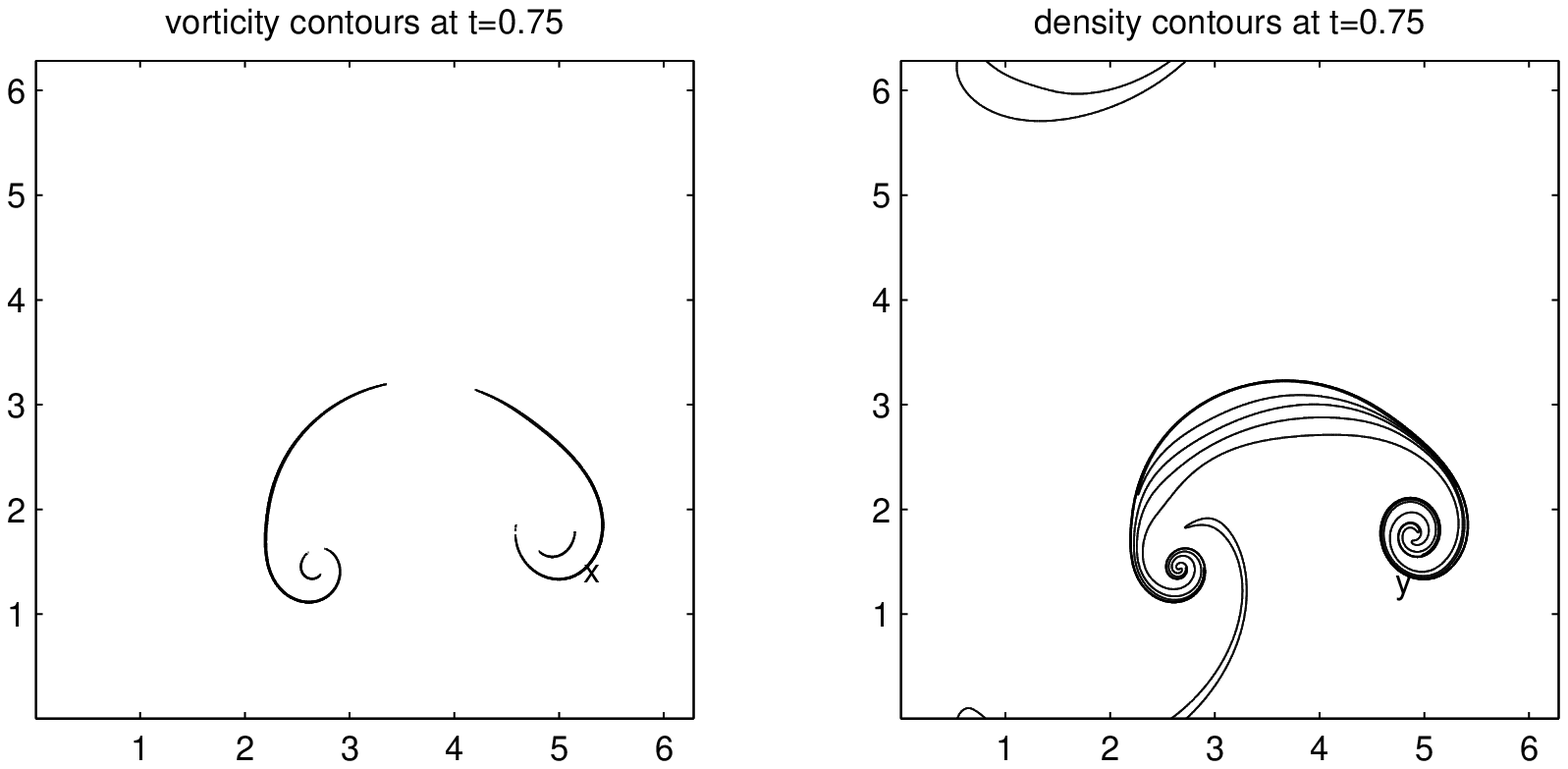}}
\end{minipage}
\caption{\label{fig:contour}Contour plots of temperature and vorticity with the resolution of $8192^2$. Here, ``x''
indicates the location of $\omega_{max}$, and ``y'' the location of $|\nabla \theta|_{max}$.}
\end{figure}

\begin{figure}
\begin{minipage}[c]{.49 \linewidth}
\scalebox{1}[1.1]{\includegraphics[width=\linewidth]{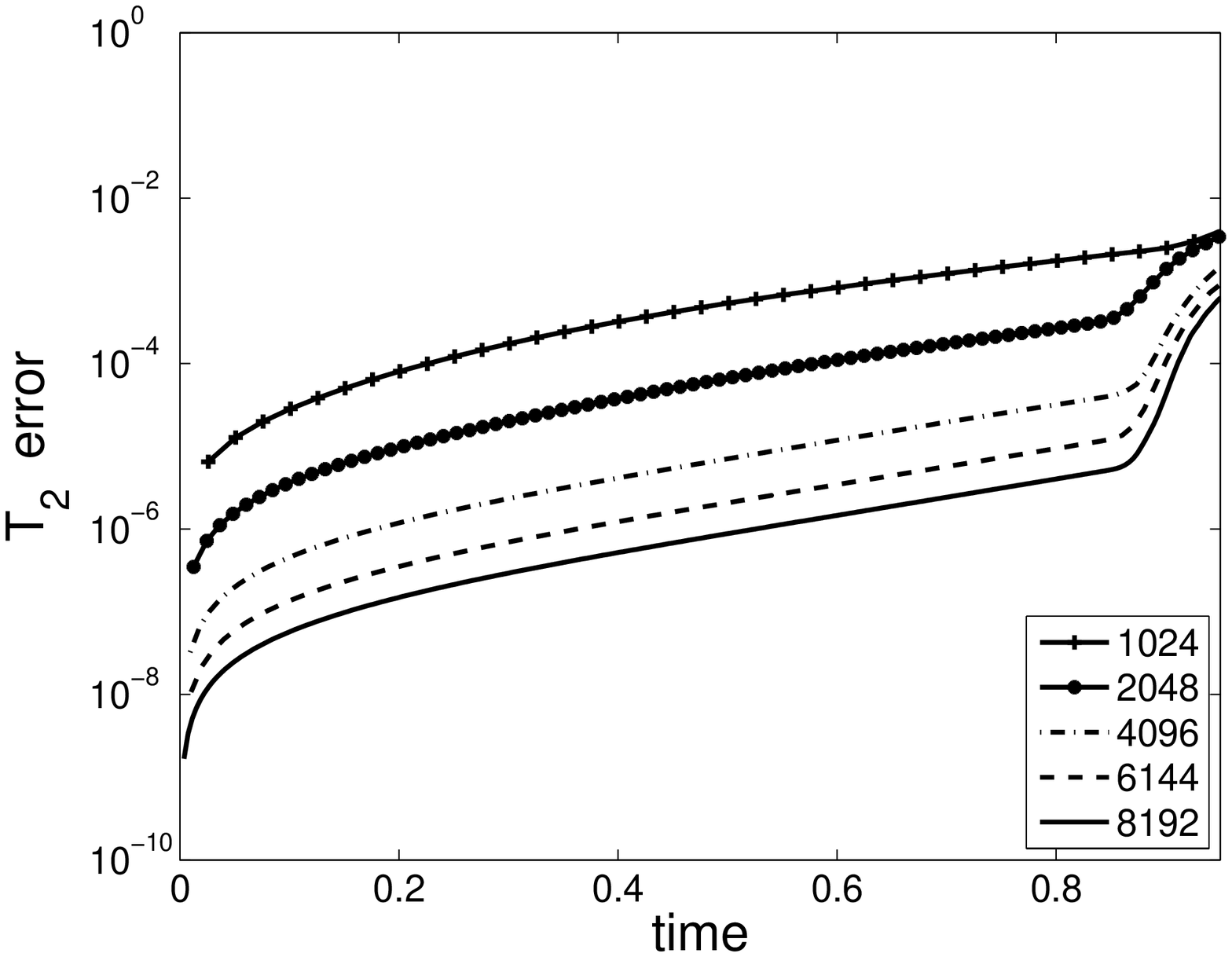}}
\end{minipage}
\begin{minipage}[c]{.49 \linewidth}
\scalebox{1}[1.1]{\includegraphics[width=\linewidth]{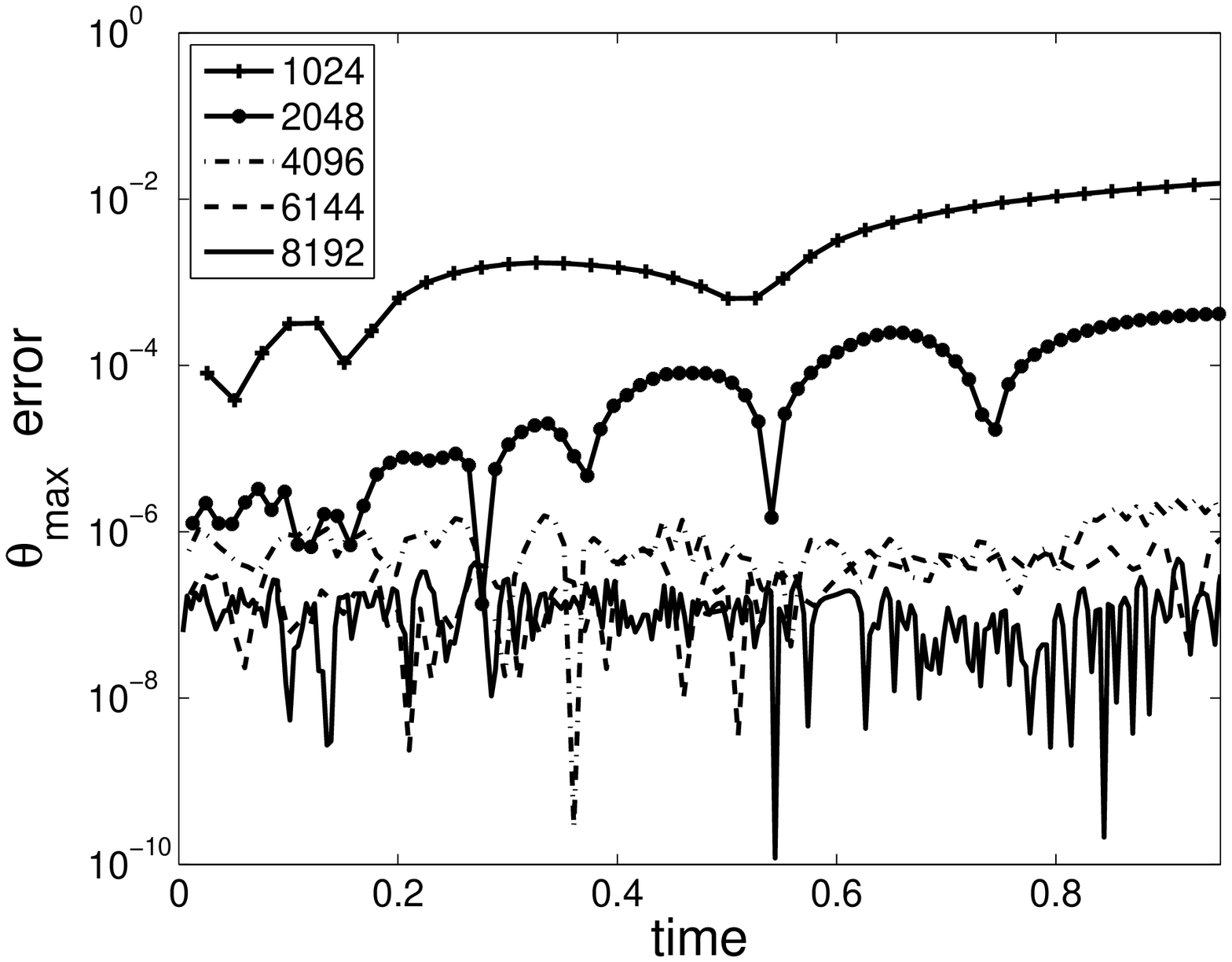}}
\end{minipage}
\begin{minipage}[c]{.49 \linewidth}
\scalebox{1}[1.1]{\includegraphics[width=\linewidth]{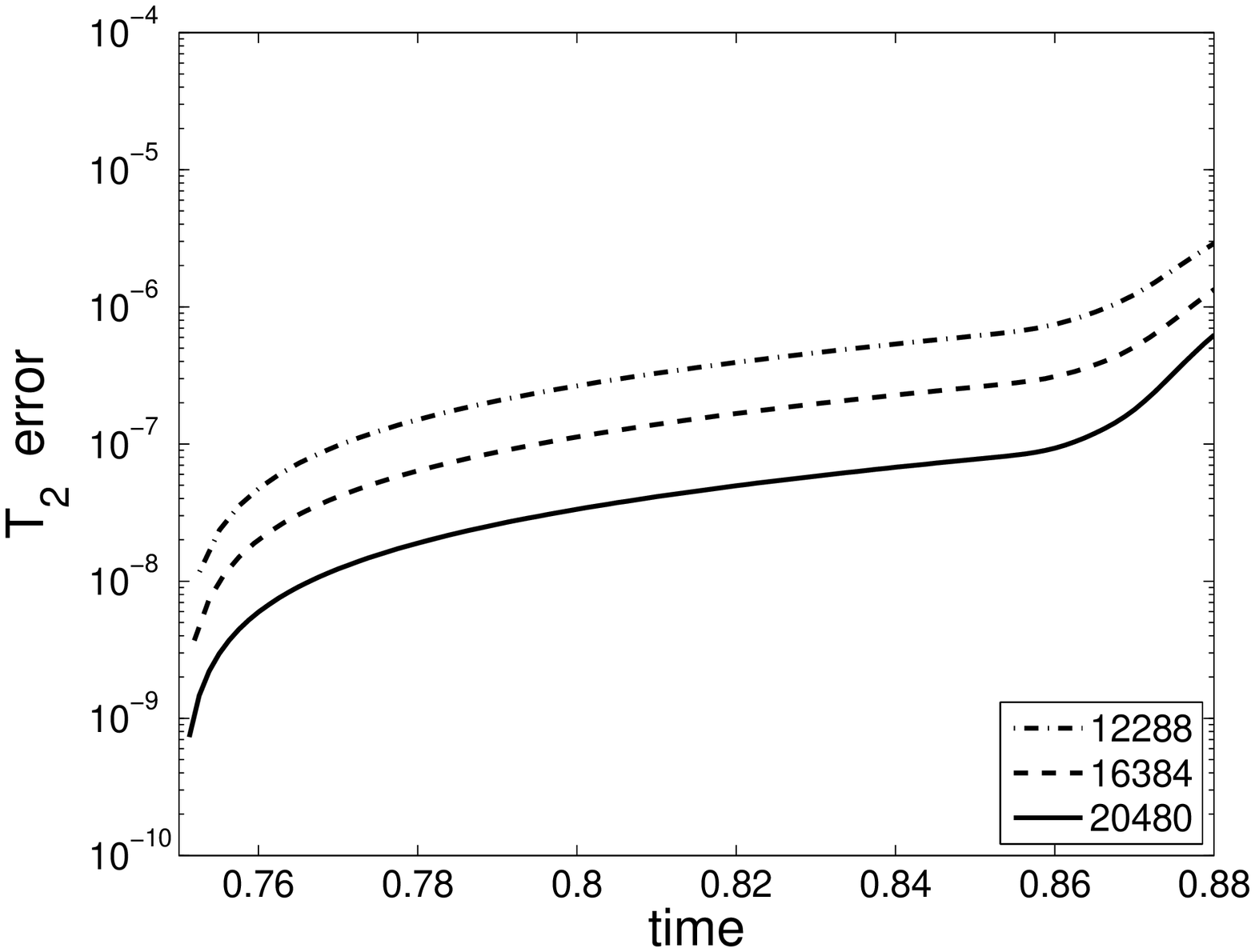}}
\end{minipage}
\begin{minipage}[c]{.49 \linewidth}
\scalebox{1}[1.1]{\includegraphics[width=\linewidth]{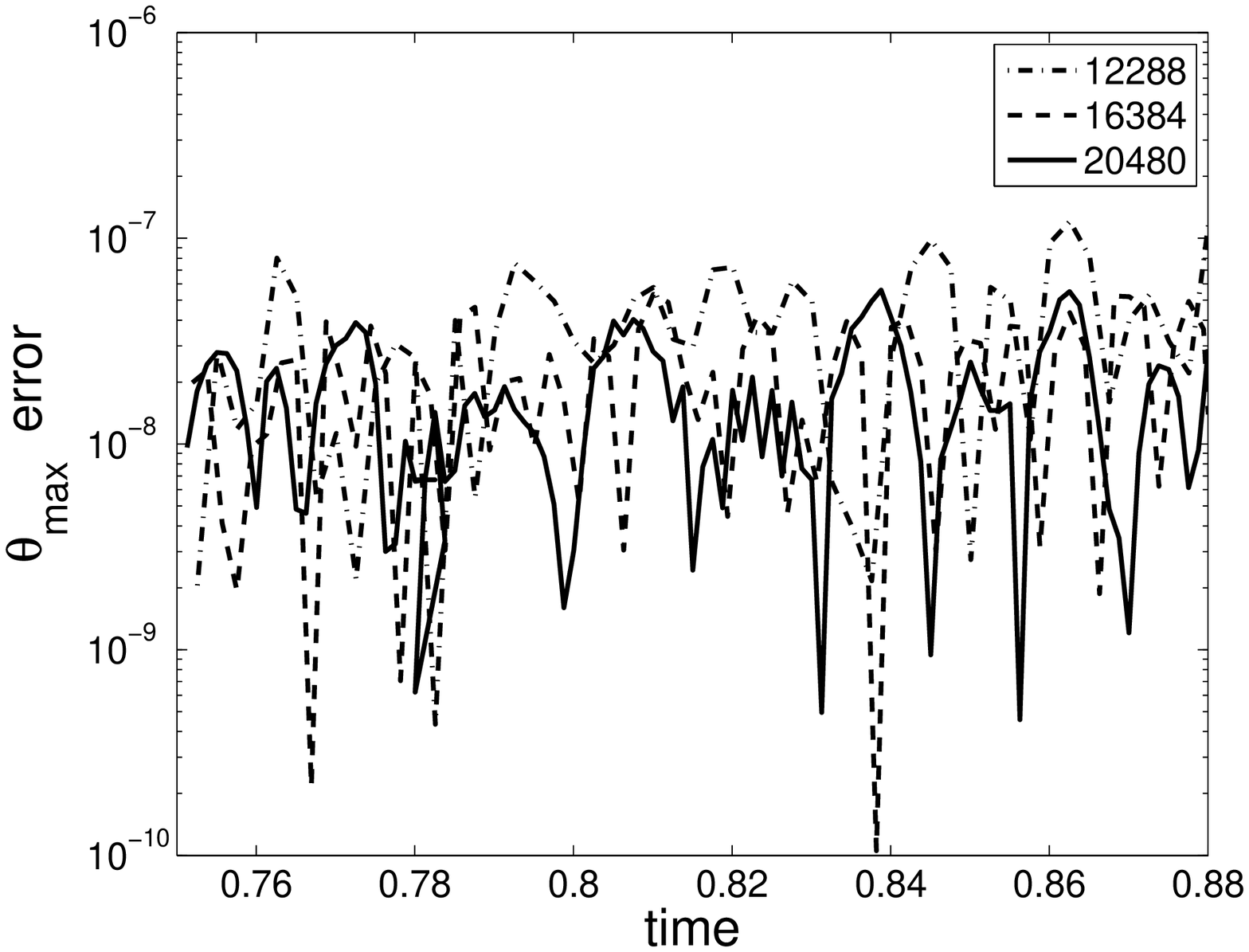}}
\end{minipage}
\begin{minipage}[c]{.49 \linewidth}
\scalebox{1}[1.1]{\includegraphics[width=\linewidth]{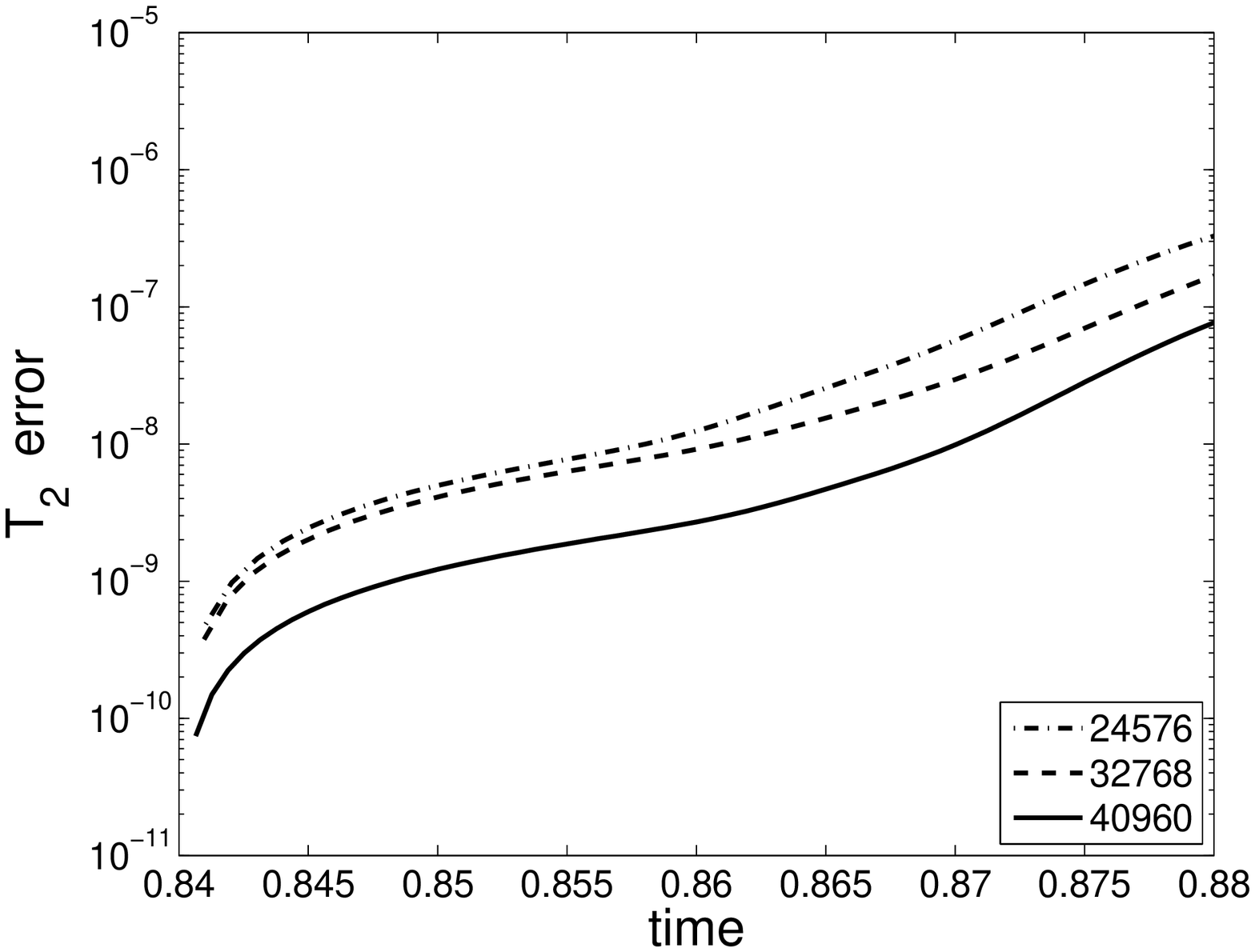}}
\end{minipage}
\begin{minipage}[c]{.49 \linewidth}
\scalebox{1}[1.1]{\includegraphics[width=\linewidth]{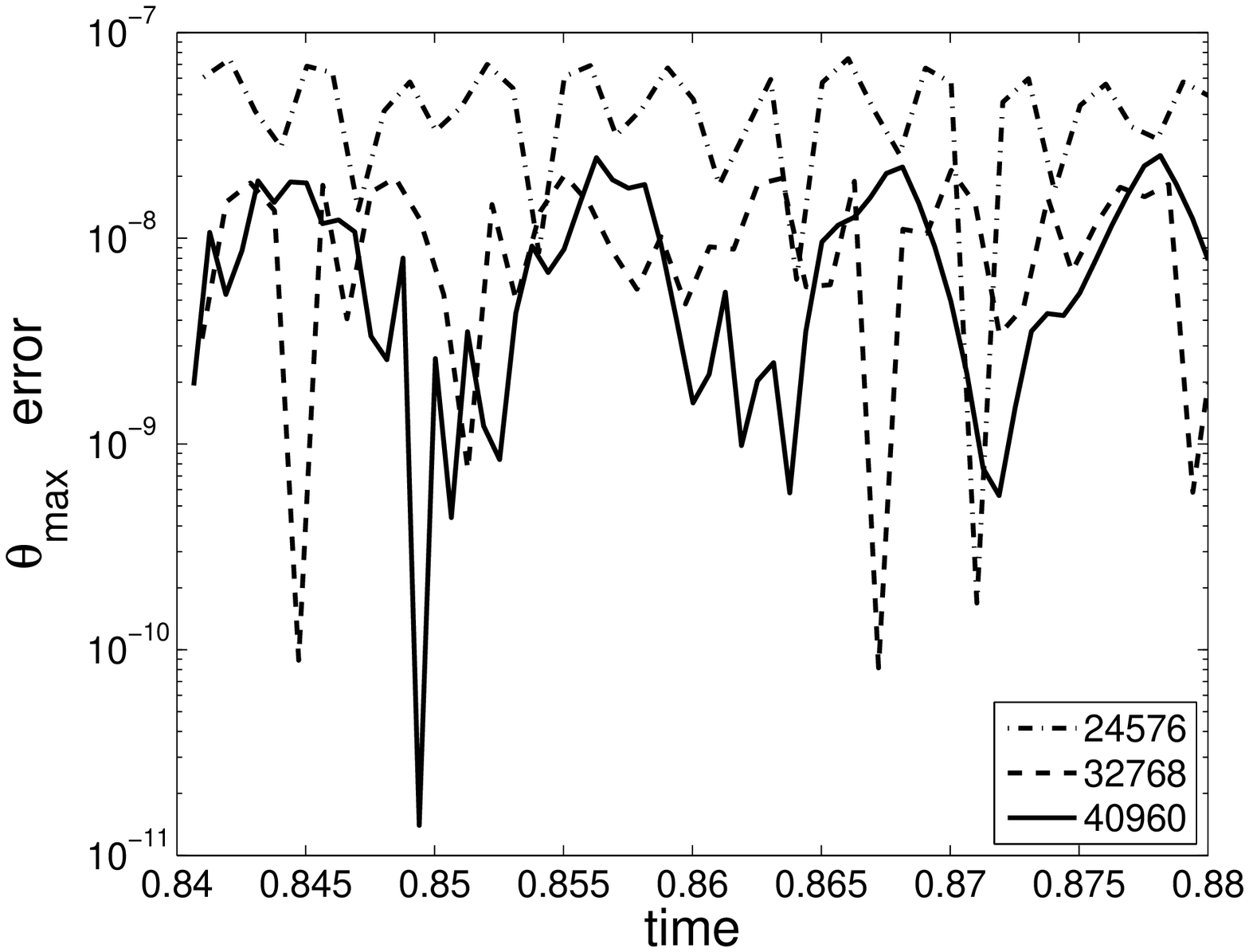}}
\end{minipage}
\caption{\label{fig:error} The evolution of the $T_2$ and maximum $\theta$ errors for all resolutions. The errors are
defined as ${(T_2(t_0)-T_2(t))}/{T_2(t_0)}$ and $|\theta_{max}(t_0) - \theta_{max}(t)|/|\theta_{max}(t_0)|$. Note that
$t_0=0$ for Stage1, $t_0=0.75$ for Stage2, and  $t_0=0.84$ for Stage3.}
\end{figure}

\begin{figure*}
\begin{minipage}[c]{.31 \linewidth}
\scalebox{1}[1]{\includegraphics[width=\linewidth]{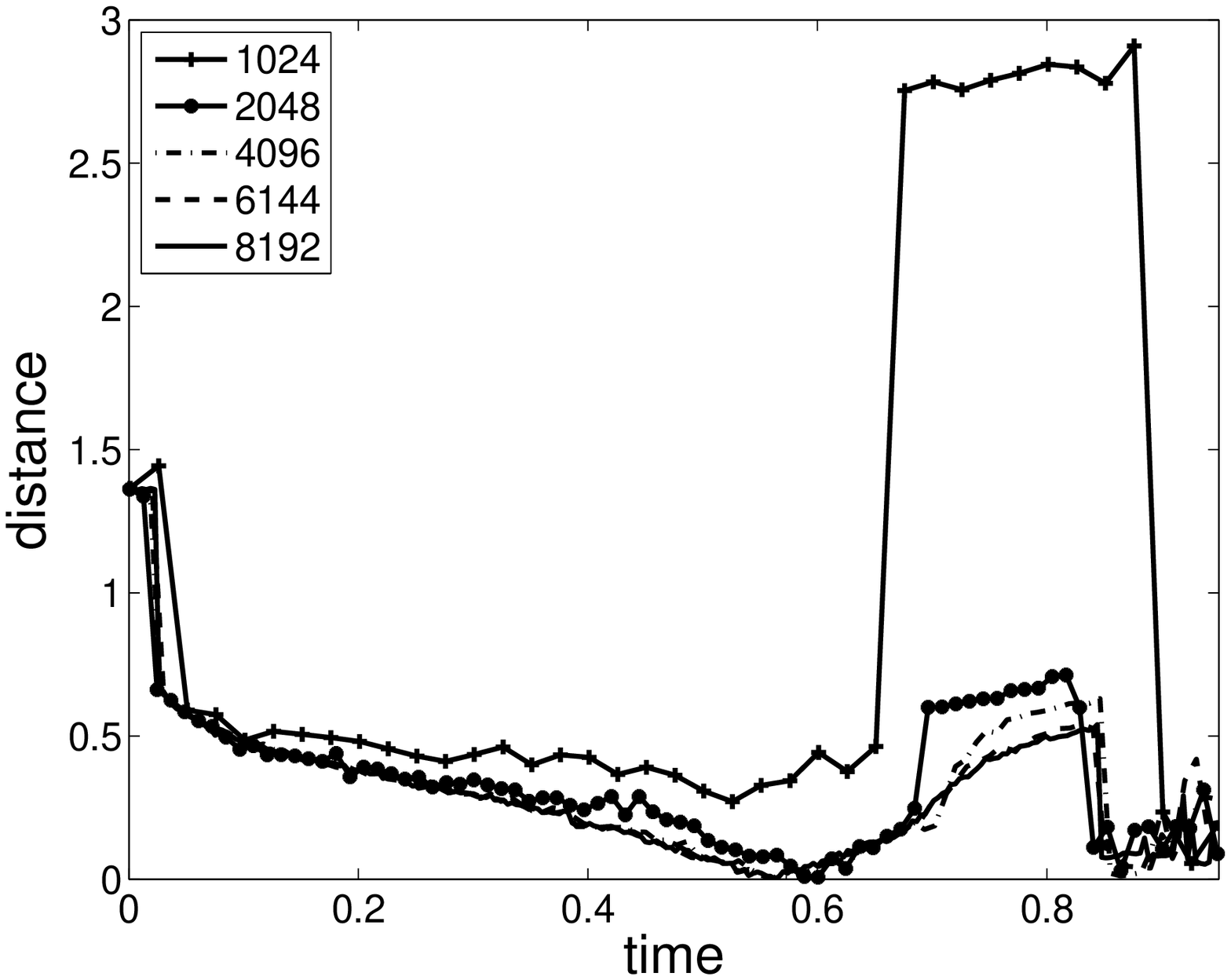}}
\end{minipage}
\begin{minipage}[c]{.31 \linewidth}
\scalebox{1}[1]{\includegraphics[width=\linewidth]{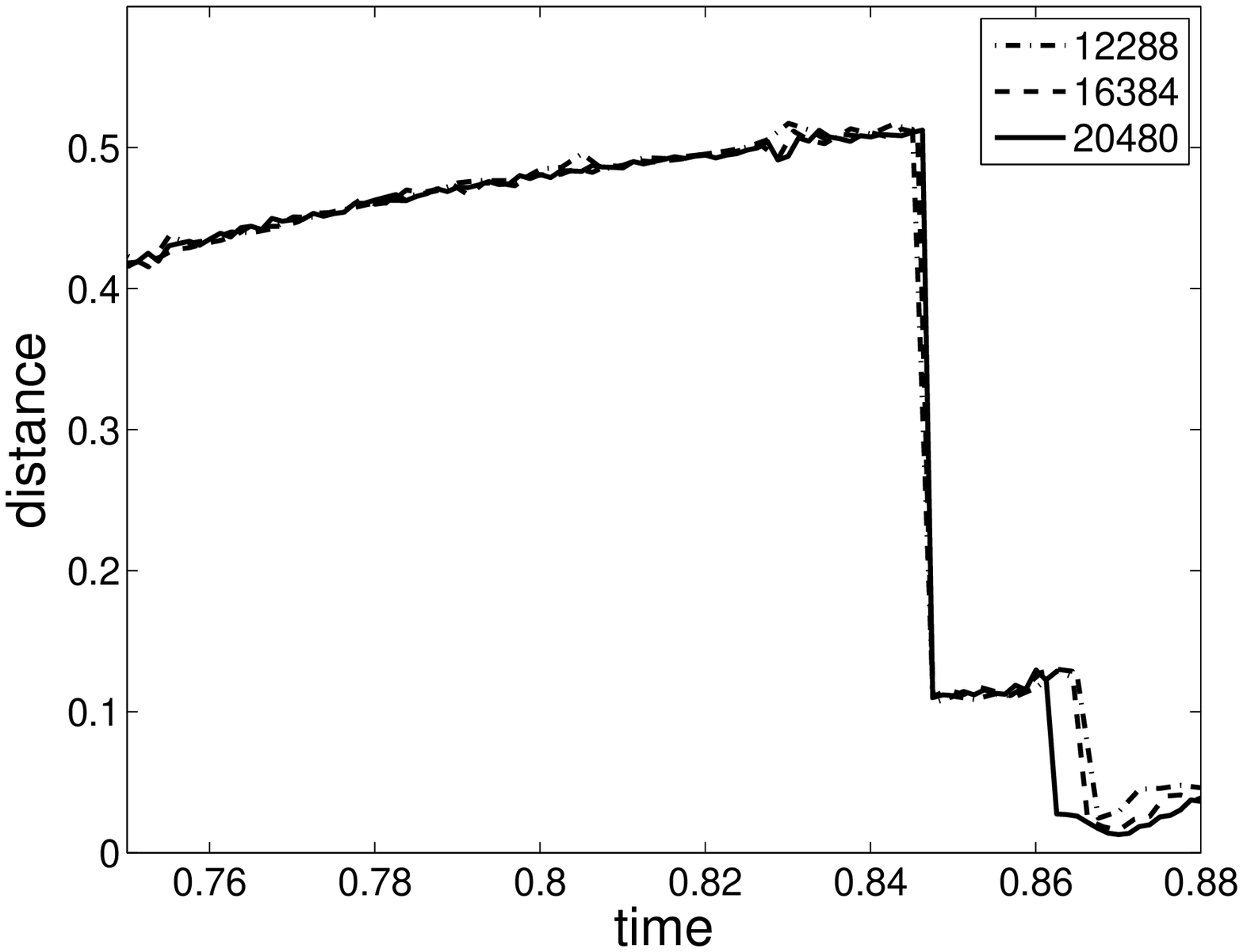}}
\end{minipage}
\begin{minipage}[c]{.31 \linewidth}
\scalebox{1}[1]{\includegraphics[width=\linewidth]{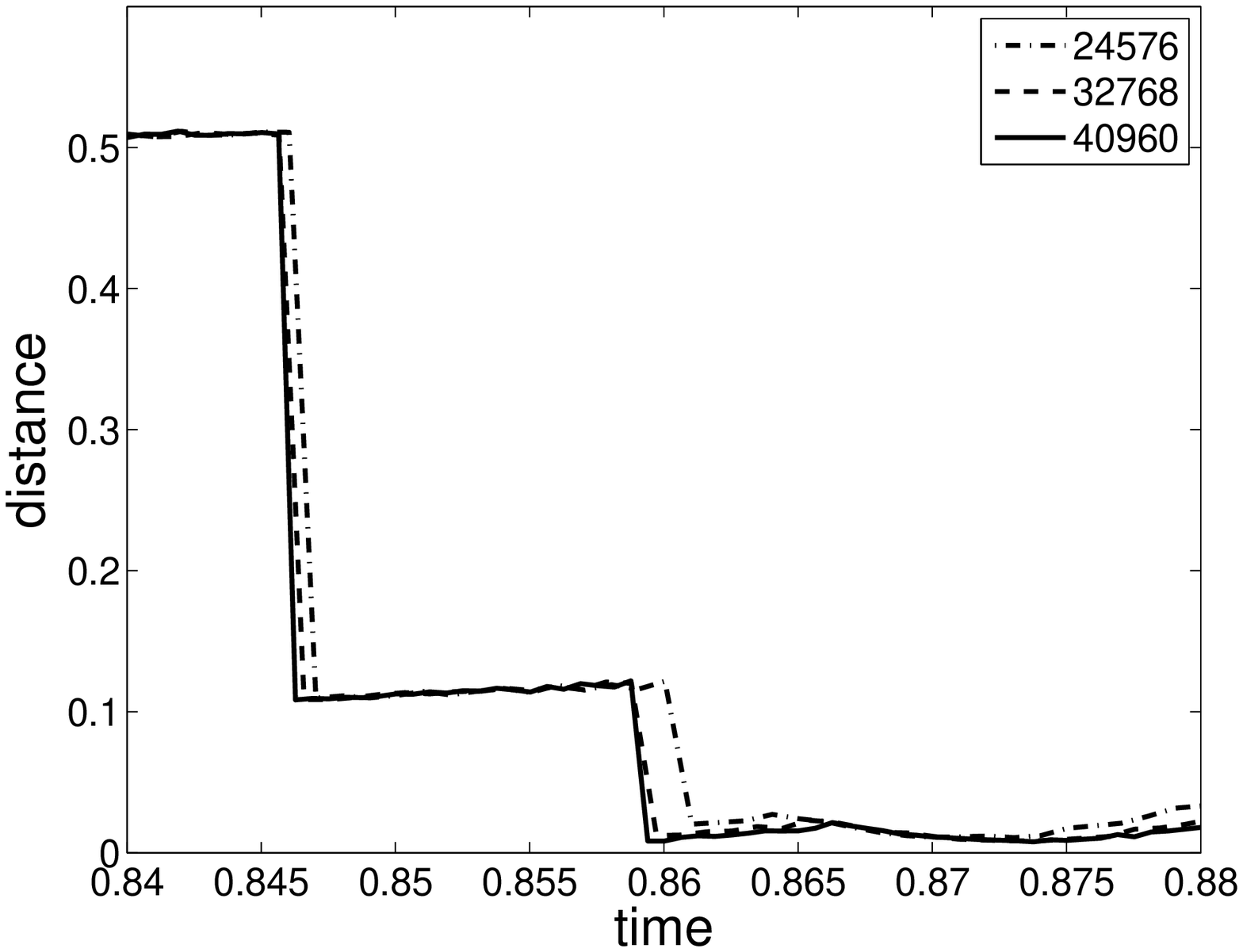}}
\end{minipage}
\caption{\label{fig:cdis} The distance between the locations of $\omega_{max}$ and $|\nabla \theta|_{max}$ at different
times.}
\end{figure*}

\begin{figure*}
\begin{minipage}[c]{0.99 \linewidth}
\begin{minipage}[c]{0.32 \linewidth}
\scalebox{1}[1]{\includegraphics[width=\linewidth]{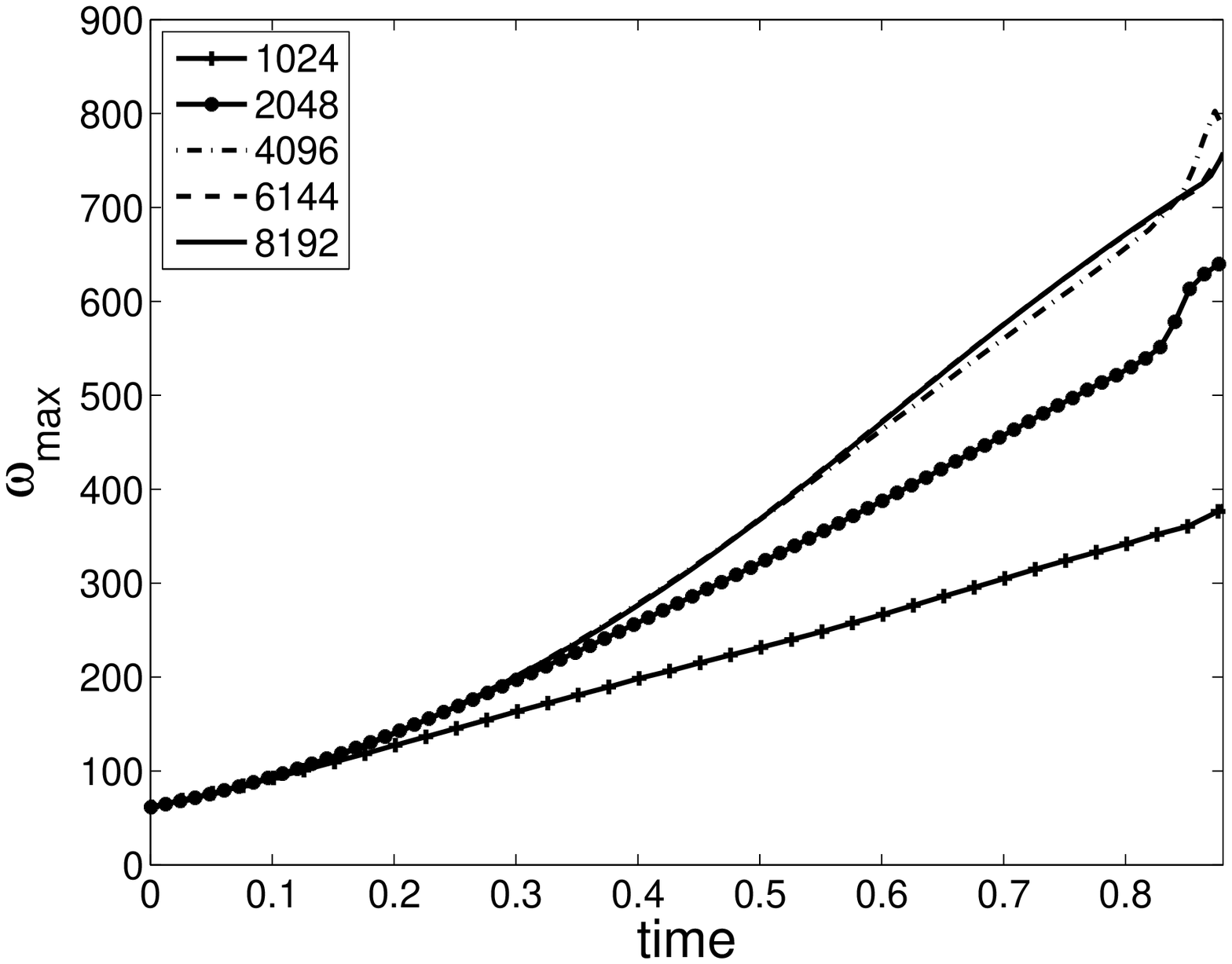}}
\end{minipage}
\begin{minipage}[c]{0.32 \linewidth}
\scalebox{1}[1]{\includegraphics[width=\linewidth]{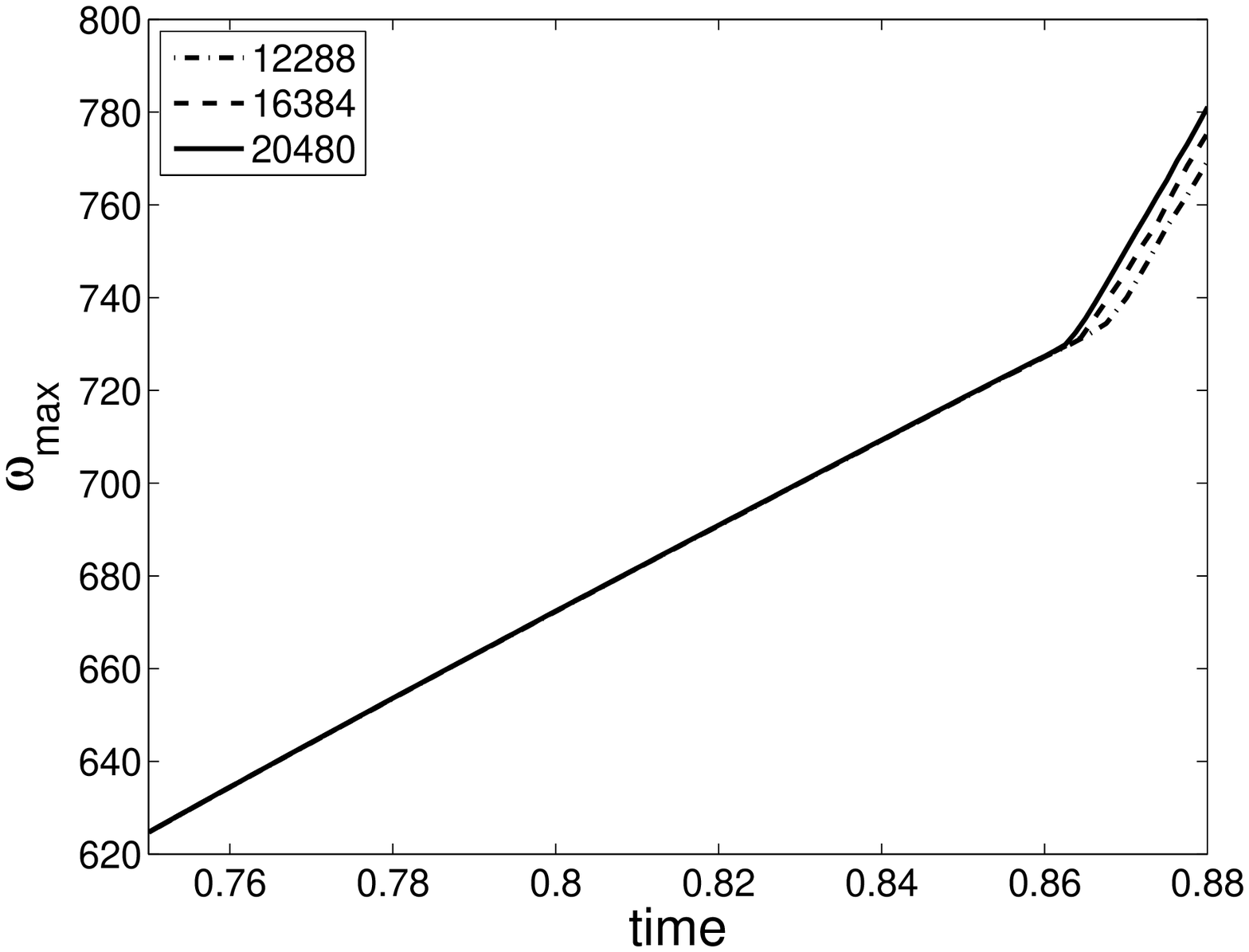}}
\end{minipage}
\begin{minipage}[c]{0.32 \linewidth}
\scalebox{1}[1]{\includegraphics[width=\linewidth]{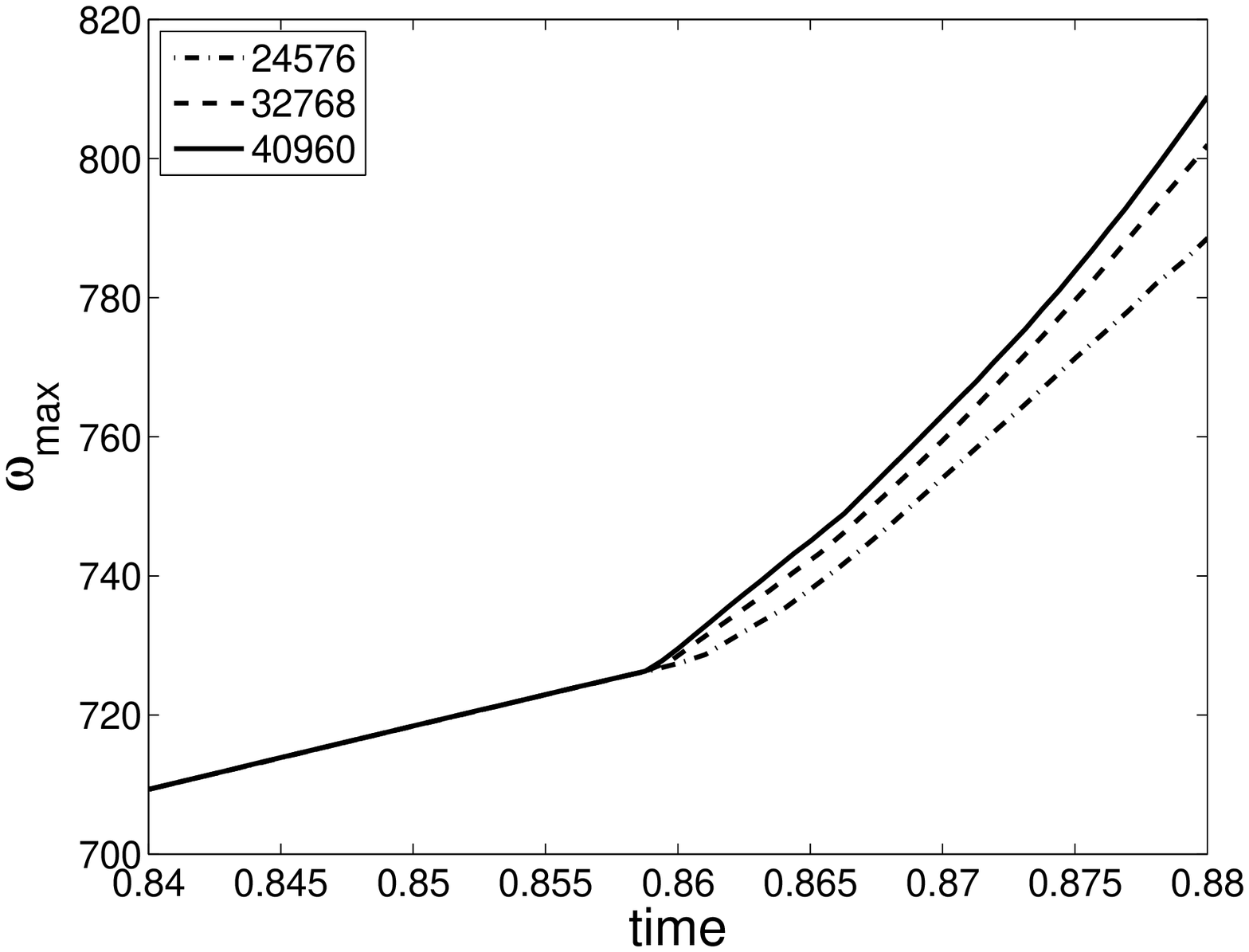}}
\end{minipage}
\caption{\label{fig:omaxsum}Time evolutions of maximum $\omega$.}
\end{minipage}
\begin{minipage}[c]{0.99 \linewidth}
\begin{minipage}[c]{0.32 \linewidth}
\scalebox{1}[1]{\includegraphics[width=\linewidth]{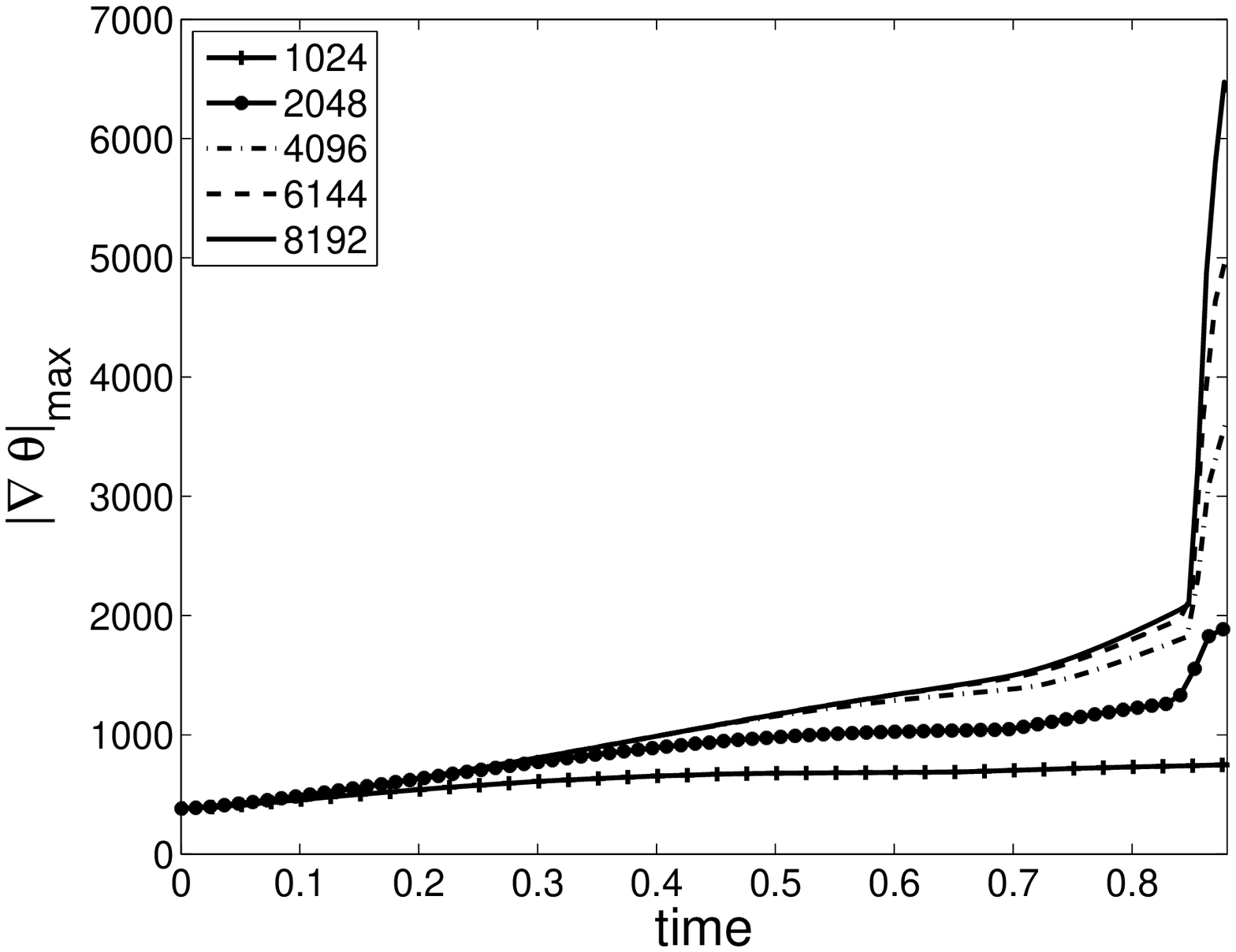}}
\end{minipage}
\begin{minipage}[c]{0.32 \linewidth}
\scalebox{1}[1]{\includegraphics[width=\linewidth]{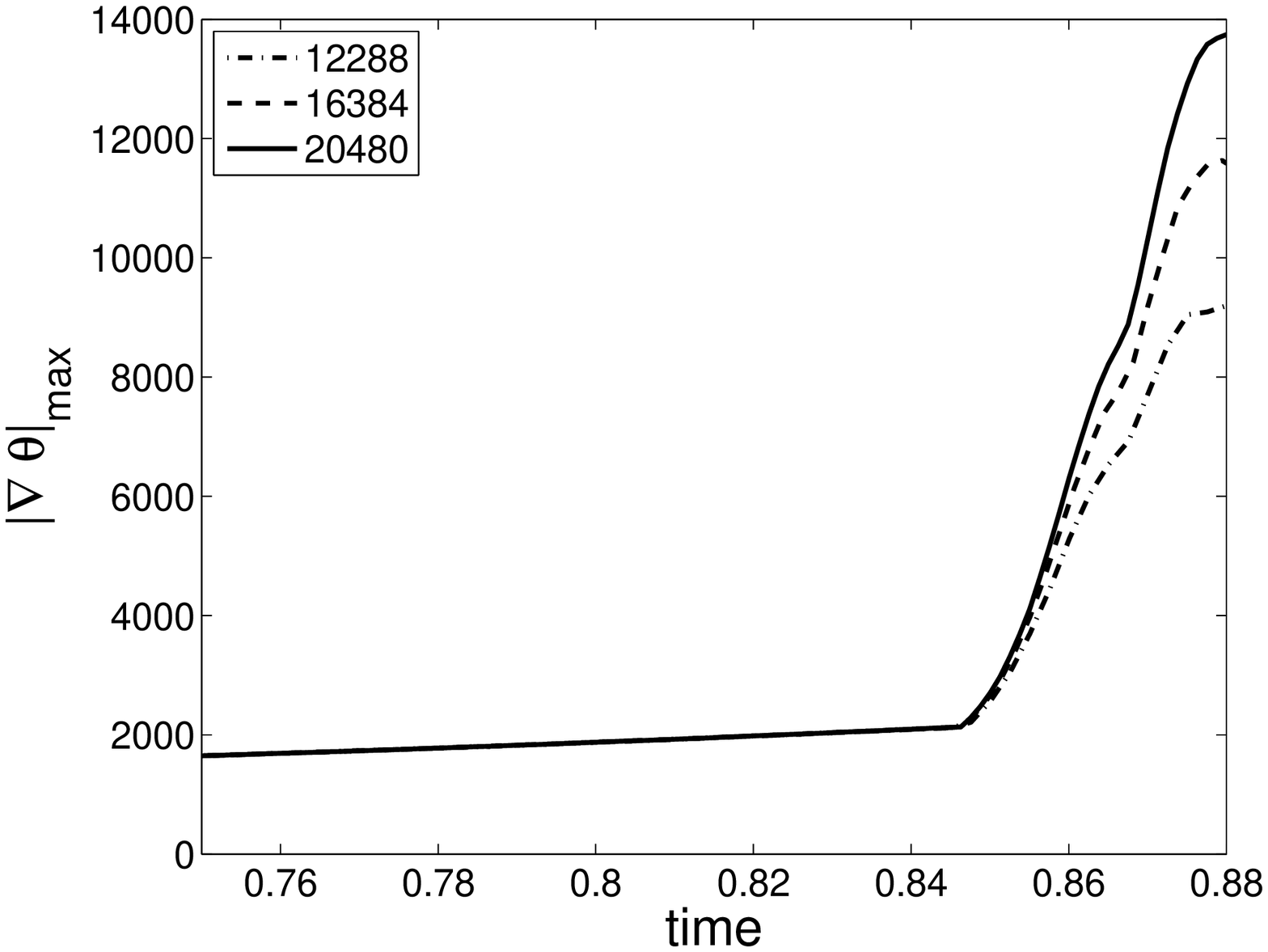}}
\end{minipage}
\begin{minipage}[c]{0.32 \linewidth}
\scalebox{1}[1]{\includegraphics[width=\linewidth]{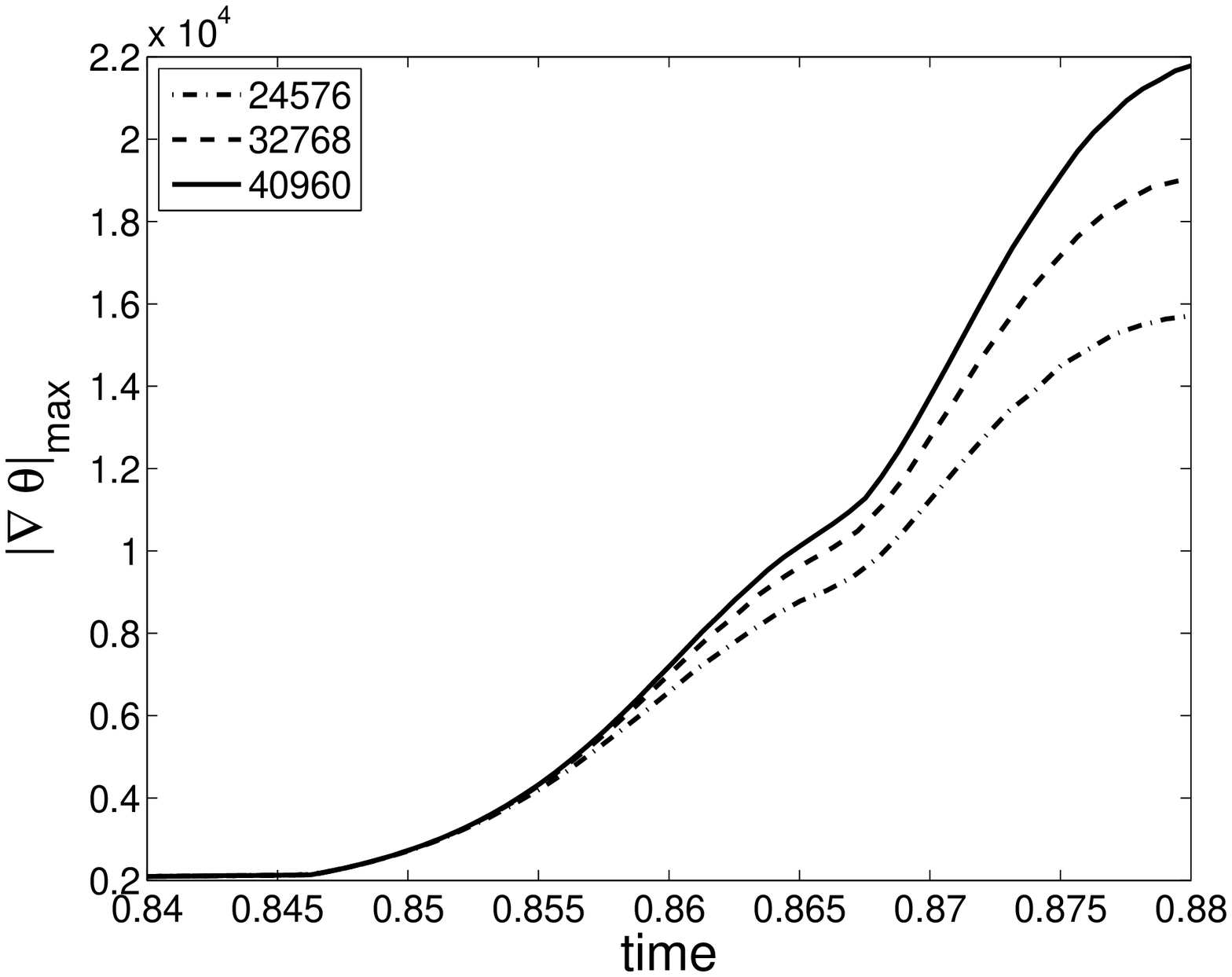}}
\end{minipage}
\caption{\label{fig:pmaxsum}Time evolutions of maximum $|\nabla \theta|$. }
\end{minipage}
\end{figure*}

\begin{description}
\item[Stage1] Full-time simulations are carried out for five resolutions: $1024^2$, $2048^2$, $4096^2$, $6144^2$,
      and $8192^2$. The time steps are $1.0 \times 10^{-3}$, $6.0 \times 10^{-4}$, $3.0 \times 10^{-4}$,
      $2.0 \times 10^{-4}$, and $1.5 \times 10^{-4}$, respectively, given by the CFL condition. For the
      $1024^2$ run, the first 1024 modes of all 2048 modes in both directions are adopted for initial data.

      For grids finer than $2048^2$, the lower modes use the $2048^2$ initial values, and higher modes are all zero
      (Stage2\&3 adopt the similar strategy).
\item[Stage2] The intermediate results at $t=0.75$ of the $8192^2$ run are used as the starting point for three resolutions:
      $12288^2$, $16384^2$, and $20480^2$. The time steps are $1.0 \times 10^{-4}$, $7.5 \times 10^{-5}$, and
      $5.0 \times 10^{-5}$, respectively.
\item[Stage3] The intermediate results at $t=0.84$ of the $20480^2$ run are used as the starting point for three resolutions:
      $24576^2$, $32768^2$, and $40960^2$. The time steps are $4.0 \times 10^{-5}$, $3.75 \times 10^{-5}$, and
      $2.5 \times 10^{-5}$, respectively.
\end{description}

In time-evolution plots of $\omega_{max}$ and $|\nabla \theta|_{max}$ (Figs. (\ref{fig:omaxsum})(\ref{fig:pmaxsum})),
the overlap between $6144^2$ and $8192^2$ curves lasts after $t = 0.8$, which makes us believe it is safe to use the
$t=0.75$ results of $8192^2$ as a starting point for Stage2. The starting point for Stage3 is decided similarly. All
simulations are carried out after the possible blowup time with our MPI C++ solver ~\cite{yin05}.

To demonstrate the accuracy of our numerical results, we will check two values which should be time independent during
the simulations due to the divergence-free constraint, the doubly-periodic condition and the inviscid transport
equation (Eq. (\ref{eq23})):
\begin{itemize}
\item
$T_2(t) = \int^{2\pi}_{0}\int^{2\pi}_{0} \theta^2(x,y,t) dxdy$,
\item
$\theta_{max}(t)$.
\end{itemize}
Fig. \ref{fig:error} shows that these global average quantities are well conserved for all simulations, and errors are
better controlled for finer grids. For the largest resolution used in this paper ($40960^2$), the $T_2$ error is below
$1.0\times10^{-7}$, and the $\theta_{max}$ error is below $3.0\times10^{-8}$ until the supposed blowup time is closing.
The $1024^2$ run has a very poor performance according to Fig. \ref{fig:error}. We show it here mainly because it is
the largest resolution currently adopted in those full 3D Euler investigations on this issue.

If a singularity is about to form at $t = T_c$ on the point $(x_c,y_c)$, the locations of $\omega_{max}$ and $|\nabla
\theta|_{max}$ should approach $(x_c,y_c)$ when $t \to T_c$, or, the distance of the locations of these two peak values
should go to zero as $t \to  T_c$. According to Fig. \ref{fig:cdis}, the distances between these two peak values in
$2048^2$, $4096^2$, $6144^2$, and $8192^2$ runs experience similar drop-down and increasing process until $t \approx
0.846$ when they suddenly drop down to 0.1. The simulations in Stage2\&3 reveal some more details: it seems that there
is another drop-down to almost zero at $t \approx 0.86$, it seems that there might be a singularity forming around that
time, and we will focus on this point in the following.

Although we eliminate the high order mode disturbance by setting all of them to zero in this study, time evolutions of
maximum $\omega$ in Stage1 are still quite similar to previous investigations~\cite{yin2006,yin2005} (Figs.
\ref{fig:omaxsum}). During the process of increasing resolutions, the $\omega_{max}$ values at the late time are always
getting higher. The higher resolutions we adopt, the larger parts of the $\omega_{max}$ evolution curves of two
neighboring resolutions will overlap. For example, the overlap between $1024^2$ and $2048^2$ lasts until $t = 0.18$,
while that of $2048^2$ and $4096^2$ lasts until $t = 0.39$, and that of $4096^2$ and $8192^2$ lasts until $t = 0.5$.
The $\omega_{max}$ time evolutions in Stage2\&3 are distinguished from Stage1. Although higher resolutions still lead
to higher $\omega_{max}$, it only happens after $t \approx 0.86$. And just before $t \approx 0.86$, the values of
$\omega_{max}$ converge for different resolutions in Stage2\&3. So, unlike Stage1, for finer grids, the overlap parts
between two neighboring curves will not become longer in Stage2\&3.

From the discussion in the above paragraph, we realize that although our numerical scheme is globally adopted in
Computational Fluid Dynamics (CFD) with solid theoretical proof~\cite{can87}, it can not generate more accurate results
for the current problem when the resolution is larger than $12288^2$. A common knowledge in CFD field is: \textbf{If
the flow field is smooth, a proper numerical scheme with higher resolution will lead to more accurate results}.  Its
inverse negative proposition is also the truth:  \textbf{If a proper numerical scheme with higher resolution will not
lead to more accurate results, the flow field is not smooth}. It is clear that there is a singularity at $t \approx
0.86$ in our problem.

In the simulation, the values of $|\nabla \theta|$ are not filtered directly. Time evolutions of $|\nabla
\theta|_{max}$ are, by all means, the traditional ways: for larger resolutions, the $|\nabla \theta|_{max}$ values at
the late time are always higher; the higher resolution we adopt, the larger overlap between the $|\nabla \theta|_{max}$
evolution curves of neighboring resolutions is. Using the $t \in [0.84,0.859]$ values of the $40960^2$ run, we estimate
that $T_c=0.86$, $\alpha$ is slightly larger than 1, and $\beta=2.891$ (Eq. \ref{eqcc}).

\begin{figure}
\begin{minipage}[c]{0.8 \linewidth}
\scalebox{1}[1]{\includegraphics[width=\linewidth]{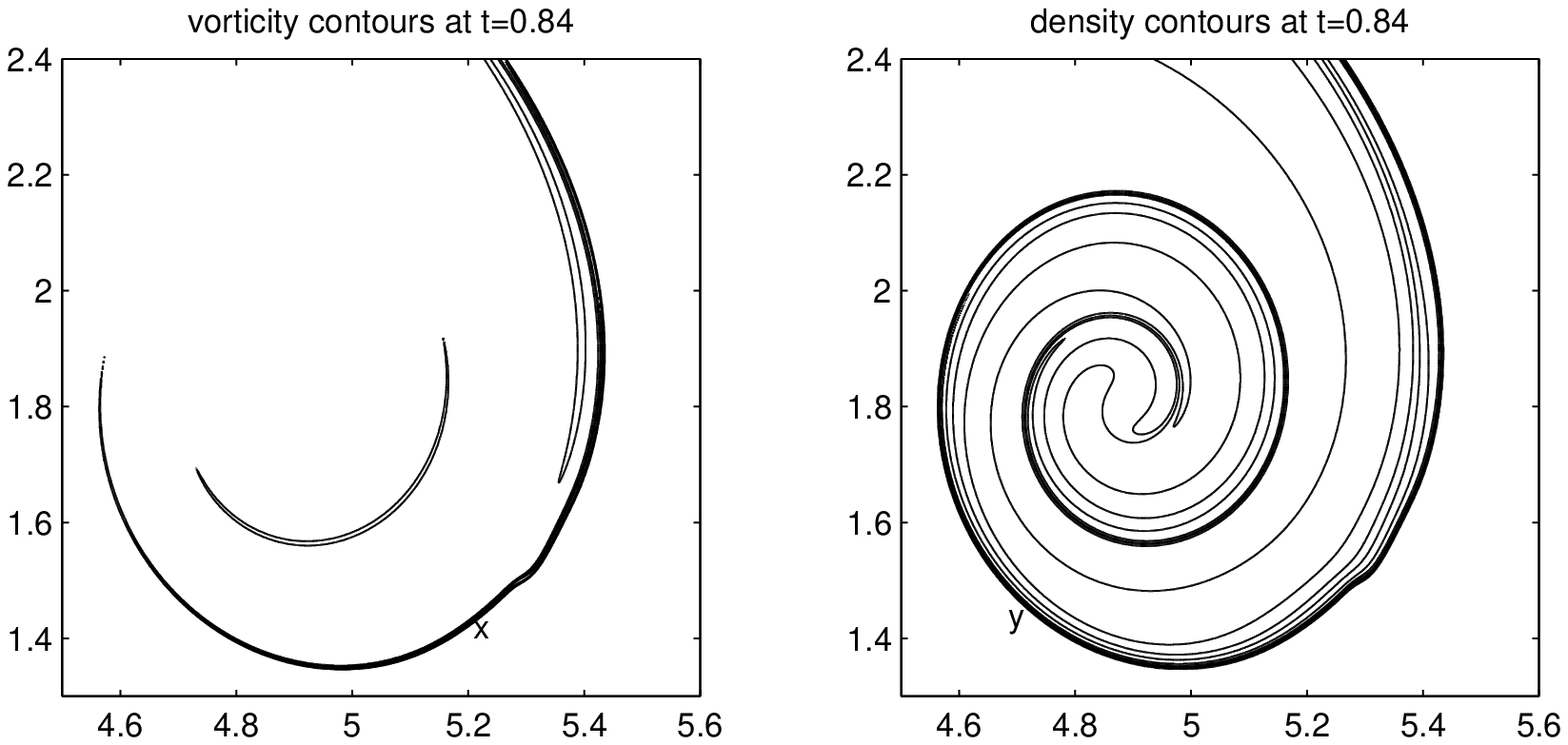}}
\end{minipage}
\begin{minipage}[c]{0.8 \linewidth}
\scalebox{1}[1]{\includegraphics[width=\linewidth]{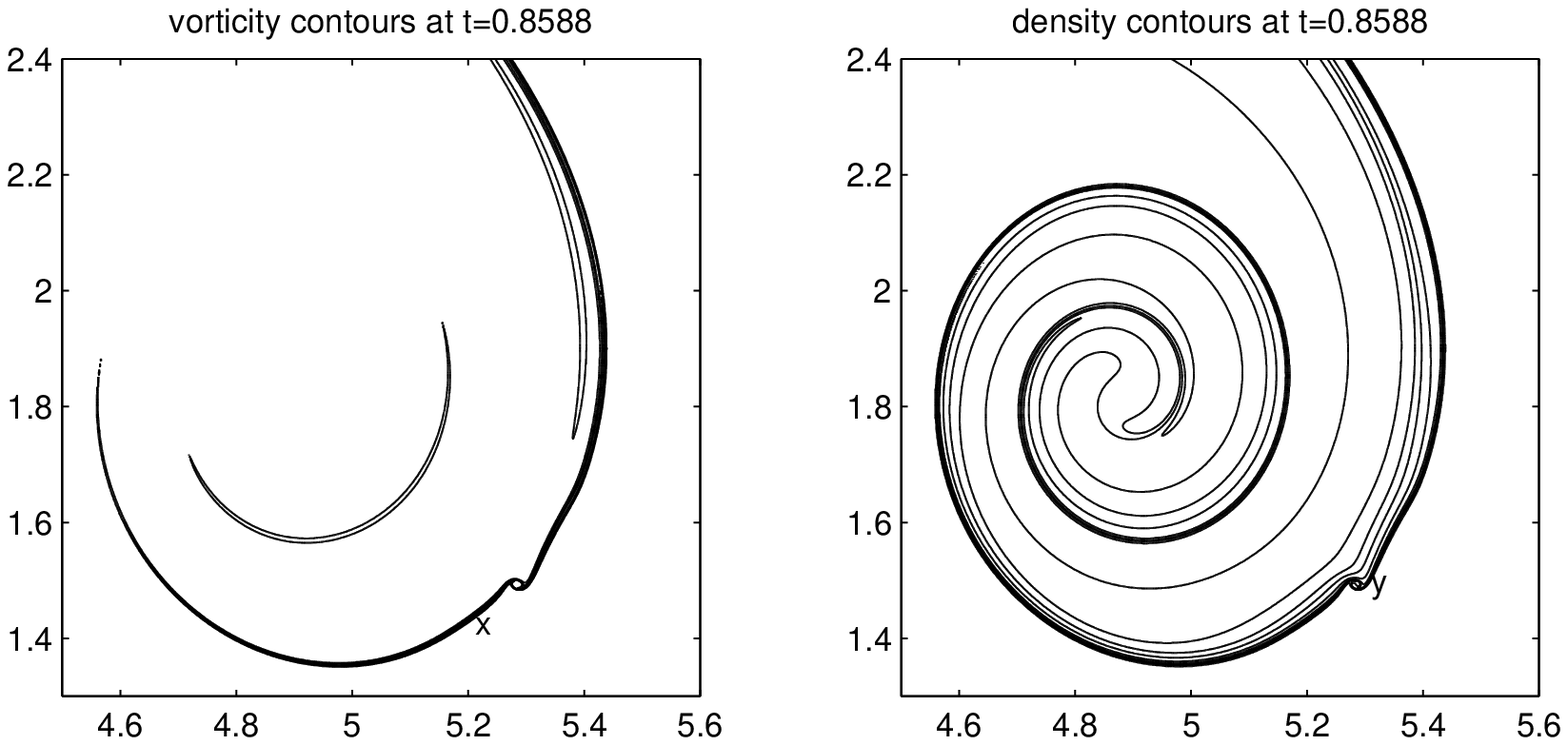}}
\end{minipage}
\begin{minipage}[c]{0.8 \linewidth}
\scalebox{1}[1]{\includegraphics[width=\linewidth]{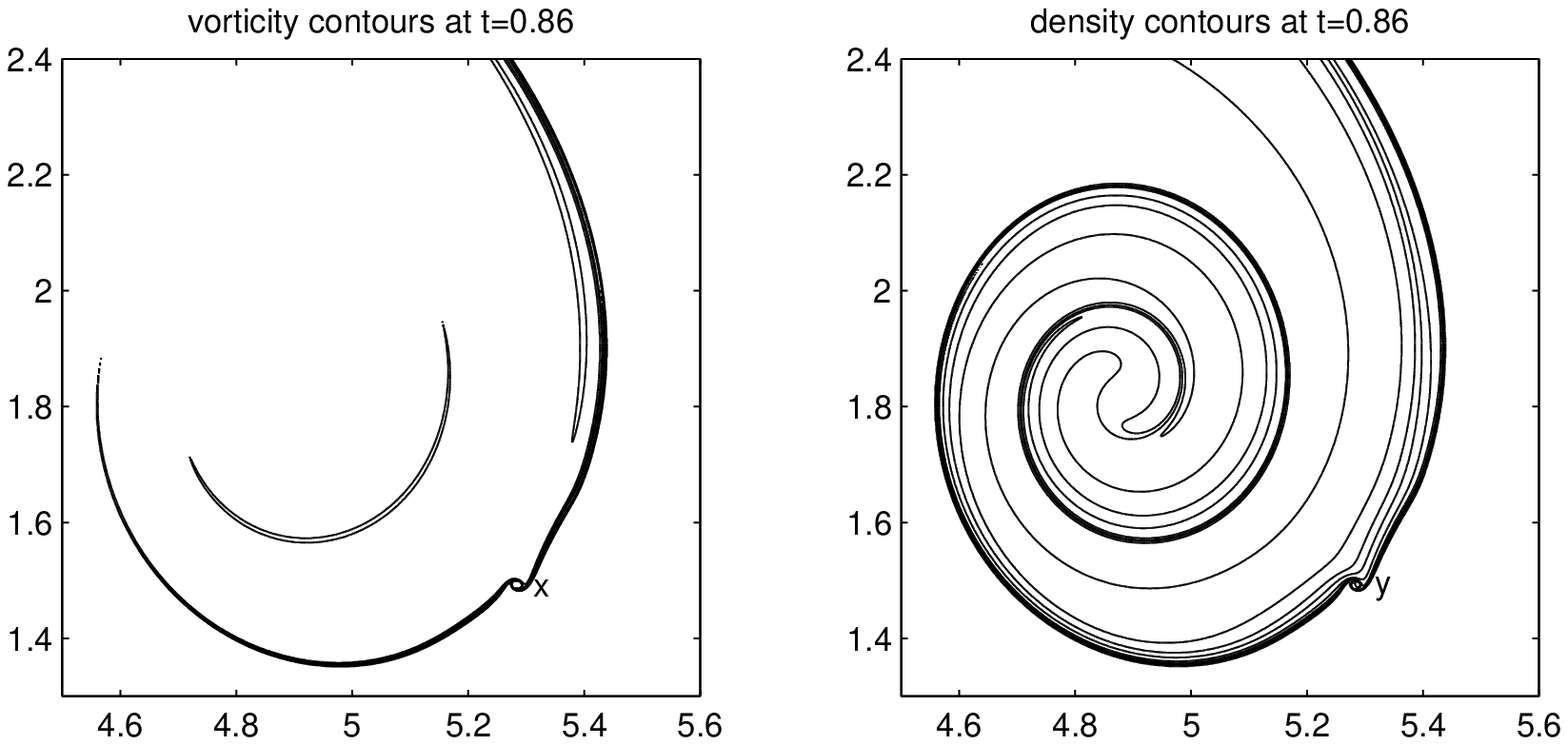}}
\end{minipage}
\caption{\label{fig:contour2}Contour plots of temperature and vorticity near the blowup time with the resolution of
$40960^2$. Here, ``x'' indicates the location of $\omega_{max}$, and ``y'' the location of $|\nabla \theta|_{max}$.}
\end{figure}

The physical process around $t = 0.86$ is shown in Fig. \ref{fig:contour2}. There is a secondary vortex forming near
[5.3, 1.5] since $t=0.84$, and it eventually becomes the strongest vortex in the whole domain at $t=0.86$, and absorb
both maximum $\omega$ and $|\nabla \theta|$ within its region.

To sum up, there are several points that make us believe that there is a singularity at $t \approx 0.86$:
\begin{itemize}
\item The distance between $\omega_{max}$ and $|\nabla \theta|_{max}$ suddenly drop down to almost zero around
      $t \approx 0.86$;
\item For enough fine resolutions, the $\omega_{max}(t)$ curves converge before $t \approx 0.86$, and they suddenly
      diverge after $t \approx 0.86$;
\item $\beta > 2$.
\end{itemize}

This work is supported by National Natural Science Foundation of China (G10502054) and the Knowledge Innovation Program
of the Chinese Academy of Sciences (Grant No.KJCX2-YW-L08). Simulations of Stage1 were finished on local Lenovo Deepcom
1800 supercomputer, and those of Stage2\&3 were carried out on Dawning 5000A (Magic cube) in Shanghai Supercomputer
Center.

\end{document}